\documentclass[preprint]{aastex} 

\usepackage{amsmath}
\usepackage{amsfonts}
\usepackage{amssymb}
\usepackage{epsfig}

\newcommand\brabar{\raisebox{-4.0pt}{\scalebox{.2}{
\textbf{(}}}\raisebox{-4.0pt}{{\_}}\raisebox{-4.0pt}{\scalebox{.2}{\textbf{)
}}}}

\begin{document}

\title{Neutrino emission from high-energy component gamma-ray bursts}

\author{Julia K.\ Becker} 
\affil{Fakult\"at f.\ Phys.\ \& Astron., Ruhr-Univ.\ Bochum, D-44780 Bochum, Germany}
\author{Francis Halzen}
\affil{Department of Physics, University of Wisconsin, Madison, WI-53706, USA}
\author{Aongus \'O Murchadha}
\affil{Department of Physics, University of Wisconsin, Madison, WI-53706, USA}
\author{Martino Olivo}
\affil{Fakult\"at f.\ Phys.\ \& Astron., Ruhr-Univ.\ Bochum, D-44780 Bochum, Germany}
\affil{Dept. of Physics and Astronomy, Uppsala University, Box 516, S-75120 Uppsala, Sweden}
\date{\today}

\begin{abstract}
Gamma-ray bursts have the potential to produce the particle energies (up to $10^{21}$\,eV) and the energy budget ($10^{44}\, \rm{erg\, yr^{-1}\, Mpc^{-3}}$) to accommodate the spectrum of the highest energy cosmic rays; on the other hand, there is no observational evidence that they accelerate hadrons. The Fermi GST recently observed two bursts that exhibit a power-law high-energy extension of the typical (Band) photon spectrum that extends to $\sim 30$ GeV. On the basis of fireball phenomenology we argue that they, along with GRB941017 observed by EGRET in 1994, show indirect evidence for considerable baryon loading. Since the detection of neutrinos is the only unambiguous way to establish that GRBs accelerate protons, we use two methods to estimate the neutrino flux produced when they interact with fireball photons to produce charged pions and neutrinos. While the number of events expected from the Fermi bursts detected to date is small, we conclude that an event like GRB941017 will be detected by the IceCube neutrino telescope if gamma-ray bursts are indeed the sources of the observed cosmic rays. 
\end{abstract}


%%%%%%%%%%%%%%%%%%%%%%%%%%%%%%%
\section{Introduction \label{introduction}}
%%%%%%%%%%%%%%%%%%%%%%%%%%%%%%%
The sources of the extragalactic cosmic rays with energies in excess
of $\sim$\,3$\times$10$^{18}$\,eV remain a mystery, but one of the
best motivated candidates is gamma-ray bursts (GRBs). Large cosmic-ray
energies can be achieved in the prompt phase of the GRB fireball where
internal shocks have the potential to accelerate charged particles up
to $\sim$10\,$^{21}$\,eV (\citealt{vietri_1995},
\citealt{waxman_1995}). Additionally, the total energy density in the
Universe of cosmic rays must be matched by a sufficiently high
hadronic energy density in the GRB with
$\rho_{\mbox{\tiny{had,\,GRB}}}\approx\rho_{\mbox{\tiny{CR}}}$. GRB
observations identify synchrotron photons produced by the electrons
accelerated in the fireball with an energy
$\epsilon_e\mbox{E}_{\mbox{\tiny{TOT}}}^{\rm{ iso}}\sim\,$10$^{53}$\,ergs,
where E$_{\mbox{\tiny{TOT}}}^{\rm{iso}}$ is the total isotropic energy
released by the burst. GRB fireballs also carry energy
$\epsilon_B\mbox{E}_{\mbox{\tiny{TOT}}}^{\rm{iso}}$ in the form of
magnetic fields, and, if they are the sources of cosmic rays, energy
$\epsilon_p\mbox{E}_{\mbox{\tiny{TOT}}}^{\rm{iso}}$ in non-thermal
protons. Further energy is carried by the thermal leptonic and hadronic parts, which is not discussed here.

GRBs emerge as credible sources for the ultra high-energy cosmic rays because their observed flux can be accommodated with an energy density in protons that is similar to that in electrons. Recent estimates of the local rate of GRBs yield a maximum of $\dot{n}_{0}\!\!\sim\!\! 1$~Gpc$^{-3}$~yr$^{-1}$ assuming that GRBs follow the star formation rate. For a stronger evolution with redshift, the local rate can be as low as $0.05$~Gpc$^{-3}$~yr$^{-1}$\,\citep{grbrate}. Using this result we estimate the electromagnetic energy density from GRBs to be in the range
$\rho_{\mbox{\tiny{em,\,GRB}}}\approx\dot{\mbox{n}}_{0}\,\epsilon_e\,\mbox{E}_{\mbox{\tiny{TOT}}}^{\rm{iso}}=\mbox{5}\times\mbox{10}^{42}\mbox{\,-\,}10^{44}\,\,\mbox{ergs\,Mpc}^{-3}\,\mbox{yr}^{-1}$. In
order for GRBs to be the sources of cosmic rays, their hadronic energy
density $\rho_{had,GRB}$ needs to produce the observed cosmic-ray
energy density
$\rho_{\mbox{\tiny{had,\,GRB}}}=\dot{\mbox{n}}_{0}\,\epsilon_p\,E_{\mbox{\tiny{TOT}}}^{\rm{iso}}\approx\,10^{44}\,\mbox{ergs\,Mpc}^{-3}\,\mbox{yr}^{-1}$. We
therefore conclude that
$\epsilon_p/\epsilon_e\approx\,1\mbox{\,-\,}20$. Discussion in, for example, \citet{WB},\,\citet{waxman2004}, and\,\citet{murase2008}, agrees that the ratio must lie around those values. As it can be seen from the rough estimate above, the actual number strongly depends on the local rate of GRBs, which is still quite uncertain. In addition, the non-thermal extragalactic spectrum is expected to extend to energies below the knee, but remains unobserved due to the larger contribution of galactic cosmic rays (see e.g. \citet{ahlers}). If this is the case, the average fraction of proton to electron energy needs to be larger.

The Fermi gamma-ray satellite observatory recently observed two
bursts, GRB090510 and GRB090902b, that show a statistically
significant deviation from the typical GRB spectrum described by the
Band function\,\citep{090510,090902b}. A flux of high energy events is
detected that extends to energies of $\sim$30\,GeV following a
power-law spectrum (Table~\ref{tab:latparms}). This is similar to the
much more luminous burst observed by EGRET in 1994,
GRB941017\,\citep{egret94}. There has been much discussion on the possible
 origin of these high-energy tails. A leptonic origin, interpreting the
high-energy component as synchrotron self Compton emission, is discussed by \citet{guetta_granot2003,peer2004,stern2004}.
Hadronic processes are also connected to the emission of high energy
photons. Proton synchrotron emission is discussed in \,\citep{dermer2, razzaque, asano}, while the production of neutral pions in
photohadronic reactions, which lead to the emission of high-energy
photons, is discussed in e.g.\ \cite{hh_04} for GRB941017. It is noted
by \citet{090510} that at least in the case of GRB090510, all
models show difficulties in explaining the emission and there is as yet no
model that perfectly fits the observation.
In this paper, we will investigate the possibility that the high
energy component results from $\pi^{0}$-decays and is therefore a
signature of proton acceleration. In particular, this model requires
relatively high baryonic loading of the jet. We will discuss this fact
in detail in this paper.

The main contribution to the initial opacity of the fireball comes from the annihilation of photons into $e^{\pm}$ pairs. The Fermi observation of a non-thermal spectrum up to an energy $E_{\rm{max}}$ of tens of GeV can be used to constrain the minimum bulk Lorentz factor $\Gamma_{\rm{min}}$ required to make the source optically thin at the time of the gamma-ray display. For all photons with energy $E\leq E_{\rm{max}}$ the condition $\tau_{\gamma\gamma}(E)<1$
must be fulfilled where $\tau_{\gamma\gamma}$ is the opacity. The observation of photons with energies of tens of GeV requires highly relativistic outflows with $\Gamma \simeq 10^3$. Because, on the other hand, the observed energy flux of order $10^{-4}\,\rm{ergs\,cm^{-2}\,s^{-1}}$ is typical of an average burst, the large boost factor implies that the photon density in the rest frame of the burst is low. This is a strong effect as the photon density is suppressed by $\Gamma^{-4}$. From $\tau_{\gamma\gamma}(E)=1$, we can determine $\epsilon_{e}$ by finding the electromagnetic energy over the volume of the fireball as a fraction of the total GRB energy. The optical depth is defined as
\begin{equation}
\tau_{\gamma\gamma}=\frac{\Delta R}{\lambda_{\gamma\gamma}}.
\end{equation}
Here, $\Delta R$ is the thickness of the fireball shell in its rest frame and $\lambda_{\gamma\gamma}$ is the photon mean free path. From the definition of the mean free path and the photon number density $n_{\gamma}$ as given e.g. in \citet{guetta}, we then obtain
\begin{equation}\label{eq:tau}
\tau_{\gamma\gamma}=\Delta R\, \sigma_{\gamma\gamma}\,n_{\gamma} =  \Delta R \, \sigma_{\gamma\gamma}\left( \frac{N_{\gamma}}{V_{\rm{shell}}} \right) =\Delta R \, \sigma_{\gamma\gamma} \left( \frac{L_{\gamma}}{16\pi c^{2} \delta t \Gamma^{4} \Delta R E_{\gamma}}  \right).
\end{equation}
Therefore, since the isotropic luminosity $L_{\gamma} = \epsilon_{e}\,E_{\mbox{\tiny TOT}}^{\rm{iso}}\,/T_{90}$, the condition $\tau_{\gamma\gamma}(E)=1$ gives
\begin{equation}
\label{eq:epse}
\epsilon_{e}\approx1-5 \times 10^{-2} \left( \frac{\Gamma}{300}\right)^{4} \left( \frac{\delta t}{10 \,\rm{ms}}\right) \left( \frac{E_{\gamma}}{1 \,\rm{MeV}}\right) \left( \frac{T_{90}}{100\, \rm{s}}\right) \left( \frac{10^{53}\, \rm{ergs}}{E_{\mbox{\tiny TOT}}^{\rm{iso}}}\right),
\end{equation}
where $\Gamma$ is the bulk Lorentz factor of the fireball, $\delta t$ is the variability timescale, $E_{\gamma}$ is the characteristic gamma-ray energy (which we take to be the peak energy of the event), $T_{90}$ is the duration of the burst, and $E_{\mbox{\tiny{TOT}}}^{\rm{iso}}$ is the total isotropic energy of the burst. We therefore estimate $\epsilon_{e}$ for GRB090510 to be $\sim0.05$ and for GRB090902b to be $\sim0.02$. From the low values of $\epsilon_e$ thus obtained protons seem to dominate the fireball. However, these numbers depend on having input a value for $E_{\mbox{\tiny{TOT}}}^{\rm{iso}}$, which cannot be measured, and so precise values of $\epsilon_e$ cannot be determined.

In this paper we will first discuss the properties of the bursts. We subsequently compute the neutrino flux inevitably produced when the protons interact with fireball photons. Their observation would provide incontrovertible evidence for the pionic origin of the additional high energy component in the burst and support the speculation that GRBs are the sources of the highest energy cosmic rays.

Is a kilometer-scale neutrino telescope such as IceCube sensitive enough to shed light on these questions? High energy neutrinos are produced in the fireball when protons produce pions in interactions with the photon field. Using the $\Delta$-resonance approximation,
\begin{equation}
p\,\gamma\rightarrow \Delta^{+}\rightarrow
\left\{
\begin{array}
{lll}n\,\pi^{+}&&\mbox{1/3 of the cases}\\
p\,\pi^{0}&&\mbox{2/3 of the cases}
\end{array}
\right.\,
\end{equation}
This gives pion ratios of $\pi^{+}:\pi^{0}=1:2$. The neutral pions decay as $\pi^{0}\rightarrow 2\gamma$, and the charged pions decay as $\pi^{+}\rightarrow\mu^{+}\,\nu_{\mu} \rightarrow e^{+}\,\nu_{e}\,{\nu}_{\mu}\,\bar{\nu}_{\mu}.$
Here, a single neutrino carries approximately $1/4$ of the $\pi^{+}$ energy and a photon carries $1/2$ of the $\pi^{0}$ energy. The calculation of the neutrino flux \citep{WB} has been performed in detail for the BATSE bursts~\citep{guetta, becker} with the following results: whereas an average burst produces only $\sim 10^{-2}$ neutrinos, bursts that are unusually energetic or nearby may produce an observable flux in a kilometer-scale neutrino telescope of order 10 events per year. We suggest that the power-law high-energy spectral feature can identify such bursts.

We will compute the neutrino fluxes expected in IceCube using two methods: the standard fireball model, and the bolometric method which relates the energy in neutrinos from the decay of charged pions to the observed photon energy assuming that it is of pionic origin. We will conclude that while the neutrino rates from the Fermi GST bursts are unexceptional, a burst like GRB941017 extending to tens of GeV energy will be detected by IceCube. IceCube observes cosmic neutrinos in a background of neutrinos produced in the atmosphere. Given that neutrinos of GRB origin are relatively energetic and that the direction and time of the events can be correlated to satellite alerts, the atmospheric background is suppressed and  few neutrino events may still represent a conclusive detection. We note that relating neutrinos to an observed gamma-ray flux using the $\Delta$-resonance approximation is extremely conservative in this context and the potential neutrino flux could be as much as a factor 4 greater~\citep{rachen, murase}.

%%%%%%%%%%%%%%%%%%%%%%%%%%%%%%%
\section{Detection of GRBs at high photon energies\label{extra_component}}
%%%%%%%%%%%%%%%%%%%%%%%%%%%%%%%%
The main contribution to the opacity of GRBs fireballs comes from the annihilation of pairs of photons into $e^{\pm}$ pairs. The observation of a non-thermal spectrum with maximum energy $E_{\rm{max}}$ requires the fireball to be optically thin to photons at the time of emission and therefore can be used to constrain the bulk Lorentz factor $\Gamma$. By generalizing \citet{Gmin} to account for arbitrary photon spectra and the energy dependence of the pair production cross-section, we can express the optical depth $\tau$ for the most energetic photon in the GRB fireball (with observed energy $E_{\rm{max}}$) as
\begin{equation}
\tau = \frac{4\pi d_{L}^{2}\delta t}{4\pi(\Gamma^{2}c\delta t)^{2}}\int^{1}_{-1}\rm{d(cos\,\theta)}\,\frac{(1-\rm{cos\,\theta})}{2} \int_{0}^{E_{\rm{max}}} \!\! dE_{\gamma}\,\sigma_{\gamma\gamma}\left(\frac{(1+z)E_{\gamma}}{\Gamma}, \,\frac{(1+z)E_{\rm{max}}}{\Gamma},\, \rm{cos\,\theta}\right)\,\frac{dN_{\gamma}}{dE_{\gamma}}(E_{\gamma}),
\label{eq:opacity}
\end{equation}
where $d_{L}(z)$ is the luminosity distance to the source assuming the same $\Lambda$CDM cosmological parameters used in \citet{090510}, $\sigma_{\gamma\gamma}(E_{1}, E_{2}, \rm{cos\,\theta})$ is the cross-section for two real photons colliding at an angle $\theta$ with energies $E_{1}$ and $E_{2}$ to produce an electron-positron pair\,\citep{gould}, and $dN_{\gamma}/dE_{\gamma}$ is the observed photon spectrum (on Earth) from the GRB. The collision kinematics must take into account the boost factor of the fireball $\Gamma$ and the fact that the observed photons have nonzero redshift $z$. The additional factors of $\Gamma$ and $(1+z)$ modifying the arguments of $\sigma_{\gamma\gamma}$ therefore transform the observed photon energies to the energies in the center-of-mass frame and at the source, respectively. We consider only the contribution to the opacity due to pair production and neglect other processes such as inverse Compton scattering and synchrotron self-absorption. The opacity due to Compton scattering is small in a GRB fireball\,\citep{razzaque2004} and synchrotron self-absorption is not important at photon energies below $\sim10^{16}$\,eV\,\citep{murase2009}. 

\begin{figure}[h!]
\centering
\includegraphics[width=0.8\textwidth]{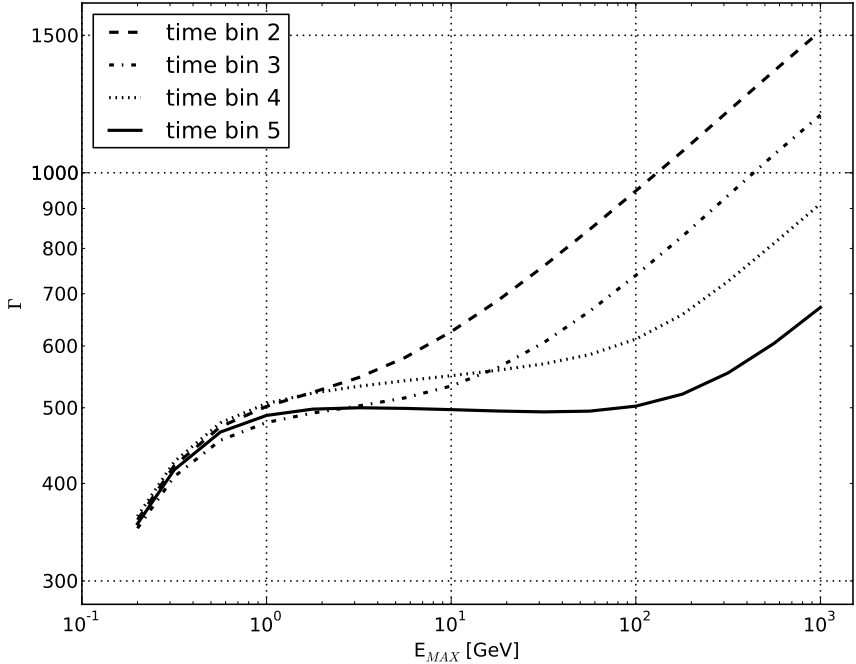}
\caption{For time bins 2-5 of GRB941017, values of $\Gamma$ such that $\tau_{\gamma\gamma}=1$ for a photon of energy $E_{\rm{max}}$ ($\Gamma_{\rm{min}}$).}
\label{fig:gVsEmax941017}
\end{figure}

The first clear detection of a GRB with a power-law component in addition to the standard Band spectral form\,\citep{band} was made by the EGRET satellite with the detection of GRB941017 with a spectrum that extended to 200 MeV, reaching the limit of EGRET's sensitivity (\citealt{egret94}). Estimates of the corresponding neutrino flux and the expected event rate in km$^{3}$-scale neutrino telescopes are presented in \citet{hh_04}. Of particular importance is the fact that since EGRET did not detect a cutoff of the power-law component, the flux can be modeled as extending to potentially very high energy $E_{\rm{max}}$. Using Eq.~\ref{eq:opacity} with the total observed photon spectrum (Band + power law), we may then find the boost factor $\Gamma$ for which $\tau=1$. We call this $\Gamma_{\rm{min}}$, the minimum boost factor for which the fireball is transparent to all photons up to energy $E_{\rm{max}}$, and is shown in Figure~\ref{fig:gVsEmax941017} for time bins 2-5 (time bin 1 showed no significant power-law component). 

The Fermi Gamma-Ray Space Telescope has been collecting very energetic photons (8\,keV\,-\,300\,GeV) from gamma-ray bursts since June 2008. As of September 2009, at the end of the first year of observations, 10 bursts have been detected by the GBM (Gamma-ray Burst Monitor) and the LAT (Large Area Telescope) simultaneously with photon energies above $\sim$100\,MeV. Fermi LAT detections have constrained the values of the jet boost factors to values of $\Gamma$ $\sim$10$^3$ for GRB080916c, GRB090510 and GRB090902b thus confirming the highly relativistic nature of the outflows. Furthermore the time integrated spectra of the recently observed GRB090510 and GRB090902b show statistically significant deviations from the Band function that extend up to $\sim$30\,GeV.

GRB090510 is a short burst ($T_{90}=2.1\mbox{\,s}$) that shows a spectral high-energy component up to $\sim$30\,GeV with a fluence of $\sim$1.8$\times$10$^{-5}\mbox{\,ergs\,cm}^{-2}$ ($\sim$\,40$\%$ of the total fluence in the energy range 10\,keV\,-\,30\,GeV)\,\citep{090510}. The isotropic energy release is estimated to be $\epsilon_{e}\,E_{\mbox{\tiny{ISO}}}\sim$1.08$\times10^{53}$\,ergs. The $\sim$31\,GeV photon is detected $\sim$1\,s after the trigger time and sets the highest lower limit on a GRB Lorentz factor ($\Gamma\sim$1200) thus proving the outflows in short gamma-ray bursts to be as highly relativistic as those in long gamma-ray bursts.

\begin{figure}[h!]
\centering
\includegraphics[width=0.8\textwidth]{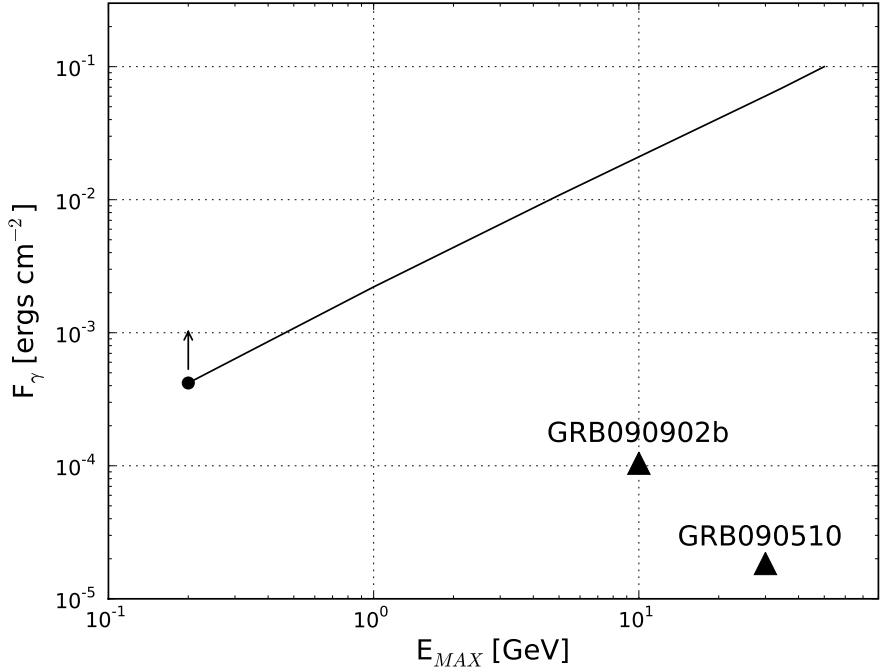}
\caption{Fluence of the high energy component above $30$\,keV as a function of the maximum energy $E_{\rm{max}}$ for GRB941017 and observed fluences for GRB090510 and GRB090902b. The arrow above the data point for GRB941017 emphasizes that EGRET was sensitive only to the part of the fluence below 200 MeV and that therefore the data point is a minimum.}
\label{fig:fluences}
\end{figure}

The extra component of the long ($T_{90}=$21.9\,s) GRB090902B extends up to $11.2$ GeV and its fluence accounts for $\sim$24$\%$ of the total fluence over the energy range 10\,keV\,-\,10\,GeV in the first 25 seconds of the prompt emission\,\citep{090902b}. The corresponding isotropic energy release is measured to be $\epsilon_{e}\,E_{\mbox{\tiny{ISO}}}\sim$3.63$\times10^{54}$\,ergs. The delay of the highest energy photon with respect to trigger time is $\sim$80\,s and the Lorentz factor of the jet is estimated to be $\Gamma\sim$1000 from opacity considerations involving the highest energy observed photon (11.16\,GeV) during the prompt emission phase. In Fig.~\ref{fig:fluences}, a comparison between the fluences of the Fermi bursts and the fluence of GRB941017 as a function of $E_{\rm{max}}$ is presented and the relative weakness of the Fermi bursts appears evident.
In Table~\ref{tab:latparms}, the spectral parameters from the Fermi GST observations of GRB090510 and GRB090902b are summarized.

\begin{table}[h!b!p!]
\caption{Spectral parameters of Fermi GRBs with power-law components. The total fluence in gamma rays $F_{\gamma}^{\mbox{\tiny{TOT}}}$ is the sum of the Band fluence $F_{\gamma}^{\mbox{\tiny{B}}}$ and the power-law fluence $F_{\gamma}^{\mbox{\tiny{HE}}}$. The fluence for GRB900510 is calculated over the energy range 10 keV--30 GeV, and the fluence for GRB09092b is calculated over the range 10 keV--10 GeV. The parameters for GRB941017 are tabulated in~\citet{egret94}.}
\begin{center}
\begin{tabular*}{0.6\textwidth}{c|cc}
           & \small{GRB090510} & \small{GRB090902b} \\[3pt] \hline \hline
\small{z}  & \small{0.903} & \small{1.822} \\
$T_{90}$   & \small{2.1\,s} & \small{21.9\,s}  \\
$F_{\gamma}^{\mbox{\tiny{TOT}}}$ & \small{$5.02\times10^{-5}$\,ergs\,cm$^{-2}$} & \small{$4.36\times10^{-4}$\,ergs\,cm$^{-2}$}  \\
$F_{\gamma}^{\mbox{\tiny{HE}}}$ & \small{$1.84\times10^{-5}$\,ergs\,cm$^{-2}$} & \small{$1.05\times10^{-4}$\,ergs\,cm$^{-2}$}  \\
$\alpha_{\gamma}$     & \small{-0.58} & \small{-0.61}  \\
$\beta_{\gamma}$      & \small{-2.83} & \small{-3.80}  \\
$\epsilon_{\gamma}$  & \small{2771\,keV} & \small{522\,keV}  \\
$\Gamma_{\gamma}$     & \small{-1.62} & \small{-1.93}  \\
$E_{\mbox{\tiny{MAX}}}$ & \small{30.53\,GeV} & \small{33.40\,GeV} \\
$\Gamma$     & \small{1260} & \small{1000}  \\
\end{tabular*}
\end{center}
\label{tab:latparms}
\end{table}

Deviations from the Band-only fit in the spectra are particularly interesting in the context of hadronic acceleration within the fireball and relate closely to predictions of neutrino fluxes detectable on Earth with km$^3$ telescopes. If these extra-components originate from $\pi^0$-\,decay photons they provide an optimal benchmark for testing models of hadronic acceleration in GRB engines.

\section{Fireball neutrinos\label{fireNus}}
In the hadronic fireball a burst of high-energy neutrinos is expected to accompany the observed prompt flux of gamma-ray photons. Assuming that electrons and protons are shock-accelerated in the same region, the neutrino spectrum can be calculated from the observed spectrum in gamma-rays using conventional fireball phenomenology as described in~\citet{guetta} and~\citet{ kappes}. For a typical GRB, the gamma-ray spectrum is usually well described by a Band function:
\begin{equation}
\label{eq:bandspec}
\frac{dN_{\gamma}}{dE_{\gamma}}=A\cdot
\begin{cases}
%\Bigg(\frac{\mbox{E}}{\mbox{100\,keV}}\Bigg)^{\alpha}\,\exp{\Bigg(-\frac{\mbox{E}}{\mbox{E}_0}\Bigg)} & \mbox{if}\quad(\alpha-\beta)\mbox{E}_0\geqslant\mbox{E} \\
%\Bigg(\frac{(\alpha-\beta)\mbox{E}_0}{\mbox{100\,keV}}\Bigg)^{\alpha-\beta}\exp(\beta-\alpha){\Bigg(\frac{\mbox{E}}{\mbox{100\,keV}}\Bigg)^{\beta}} & \mbox{if}\quad(\alpha-\beta)E_0\leqslant \mbox{E} \\
%\Bigg(\frac{{\big \epsilon}}{\rm{100\,keV}}\Bigg)^{\alpha}\,\exp{\left(-\frac{E}{E_{0}}\right)} & \rm{if}\quad(\alpha-\beta)\rm{E}_{0}\geqslant E \\
%\left(\frac{(\alpha-\beta)E_0}{\rm{100\,keV}}\right)^{\alpha-\beta}\exp(\beta-\alpha){\left(\frac{E}{\rm{100\,keV}}\right)^{\beta}} & \rm{if}\quad(\alpha-\beta)E_0\leqslant E \\

\Bigg(\frac{\mbox{E$_{\gamma}$}}{\mbox{100\,keV}}\Bigg)^{\alpha}\,\exp{\Bigg(-\frac{\mbox{E$_{\gamma}$}}{\mbox{$\epsilon$}_{\gamma}}\Bigg)} & \mbox{if}\quad(\alpha-\beta)\mbox{$\epsilon$}_{\gamma}\geqslant\mbox{E$_{\gamma}$} \\
\Bigg(\frac{\left(\mbox{$\alpha-\beta$}\right)\mbox{$\epsilon$}_{\gamma}}{\mbox{100\,keV}}\Bigg)^{\alpha-\beta}\exp(\beta-\alpha){\Bigg(\frac{\mbox{E$_{\gamma}$}}{\mbox{100\,keV}}\Bigg)^{\beta}} & \mbox{if}\quad(\alpha-\beta)\epsilon_{\gamma}\leqslant \mbox{E$_{\gamma}$} \\

\end{cases}
\end{equation}
The interaction of accelerated protons with a power-law distribution with GRB photons results in a broken power-law neutrino spectrum. The resulting neutrino spectrum is then expected to follow
\begin{equation}
\frac{dN_{\nu}}{dE_{\nu}} = f_{\nu}\cdot
\begin{cases} 
\epsilon_1^{\alpha_{\nu}-\beta_{\nu}}\rm{E}_{\nu}^{-\alpha_{\nu}}  & \mbox{if }\rm{E}_{\nu}<\epsilon_1  \\
\rm{E}_{\nu}^{-\beta_{\nu}}, & \mbox{if }\epsilon_1<\rm{E}_{\nu}<\epsilon_2 \\
\epsilon_2^{\gamma_{\nu}-\beta_{\nu}}\rm{E}_{\nu}^{-\gamma_{\nu}}  & \mbox{if }\rm{E}_{\nu}>\epsilon_2  \\
\end{cases}
\end{equation}
where the spectral indices are related to the indices of the photon spectrum according to
\begin{equation}
\alpha_{\nu}=3-\beta_{\gamma}\qquad\beta_{\nu}=3-\alpha_{\gamma}\qquad\gamma_{\nu}=\beta_{\nu}+2 .
\end{equation}
The first break energy $\epsilon_1$ is determined by the production threshold for the $\Delta$-resonance, where
\begin{equation}
\label{eq:e1b}
\epsilon_1=7.5\times10^5\,\mbox{GeV}\,\frac{1}{(1+z)^2}\,\bigg(\frac{\Gamma}{10^{2.5}}\bigg)^2\,\bigg(\frac{\mbox{MeV}}{\epsilon_{\gamma}}\bigg).
\end{equation}
$\epsilon_{\gamma}$ is the break energy of the Band function, $\Gamma$ is the jet boost factor and $z$ the redshift of the source. The spectrum steepens when pions lose energy due to synchrotron radiation prior to decay; the second break energy $\epsilon_2$ is given by
\begin{equation}
\label{eq:e2b}
\epsilon_2=10^{7}\,\mbox{GeV}\,\frac{1}{1+z}\,\sqrt{\frac{\epsilon_e}{\epsilon_B}}\,\bigg(\frac{\Gamma}{10^{2.5}}\bigg)^4\,\bigg(\frac{\delta t}{10\,\mbox{ms}}\bigg)\,\sqrt{\frac{10^{52}\,\mbox{ergs s}^{-1}}{L_{\gamma}^{\mbox{\tiny{B}}}}}.
\end{equation}
$\epsilon_e$ is the fraction of the total burst energy available to electrons, $\epsilon_B$ is the fraction of total energy going into the magnetic field $\vec{B}$ and $\delta t$ is the variability timescale. $L_{\gamma}^{\mbox{\tiny{B}}}$ is the isotropic luminosity of the Band function, given by
\begin{equation}
L_{\gamma}^{\mbox{\tiny{B}}}=\frac{E_{\gamma}^{\mbox{\tiny{B}}}}{\Delta t}=\frac{4\pi\,d_L^2(z)\,F_{\gamma}^{\mbox{\tiny{B}}}}{\Delta t(1+z)}
\end{equation}
with $\Delta t=T_{90}$ and $d_L$ the luminosity distance for $\Omega_m=0.27$, $\Omega_{\Lambda}=0.73$ and $h=0.71$. The Band gamma-ray energy fluence $F_{\gamma}^{\mbox{\tiny{B}}}$ is defined as 
\begin{equation}
\label{eq:bandfluence}
F_{\gamma}^{\mbox{\tiny{B}}} =\Delta t \int^{\infty}_{0} dE_{\gamma}\, E_{\gamma}\, \frac{dN_{\gamma}}{dE_{\gamma}}
\end{equation}
where the gamma-ray spectrum $dN_{\gamma}/dE_{\gamma}$ is the Band spectrum given in Eq.~\ref{eq:bandspec}. 

The normalization of the neutrino flux is determined by the efficiency of pion production. The relation between the gamma-ray and neutrino spectra is given by
\begin{equation}
\label{eq:fracflu}
F_{\nu}=x\cdot F_{\gamma}^{\mbox{\tiny{B}}}\qquad\mbox{where}\qquad x=\frac{1}{8}\frac{\epsilon_p}{\epsilon_e}\bigg[1-(1-\langle x_{p\rightarrow\pi}\rangle)^{\Delta R/\lambda_{p\gamma}} \bigg]
\end{equation}
and
\begin{equation}
\frac{\Delta R}{\lambda_{p\gamma}}=\bigg(\frac{L_{\gamma}^{\mbox{\tiny{B}}}}{10^{52}\mbox{\small{erg s}}^{-1}}\bigg)\bigg(\frac{10\,\mbox{ms}}{\delta t}\bigg)\bigg(\frac{10^{2.5}}{\Gamma}\bigg)^4\bigg(\frac{\mbox{MeV}}{\epsilon_{\gamma}}\bigg). 
\end{equation}
In Eq.~\ref{eq:fracflu}, $\langle x_{p\rightarrow\pi}\rangle=0.2$ is the fraction of proton energy transferred to a pion in a single interaction and $\epsilon_{\gamma}$ is the break energy of the Band function. The neutrino energy fluence $F_{\nu}$ is defined analogously to $F_{\gamma}^{\mbox{\tiny{B}}}$ (Eq.~\ref{eq:bandfluence}).

\begin{table}[h!b!p!]
{
\caption{Parameters for fireball neutrino fluxes. Typical values of the WB model are reported. Quantities marked with ` * ' are not determined for the particular burst and standard WB model values are used.}
\label{tab:nuParms}
}
{
\begin{center}
\begin{tabular*}{0.6\textwidth}{c|c|c|c}
 & \small{WB} & \small{GRB090510} & \small{GRB090902b} \\[3pt]
\hline \hline
\small{$\epsilon_{\gamma}$} & \small{0.1-1 MeV}  & \small{$2.771$ MeV} & \small{$0.522$ MeV} \\
\small{$\alpha_{\gamma}$} & \small{-1}   & \small{-0.58} & \small{-0.61}\\
\small{$\beta_{\gamma}$} & \small{-2}  & \small{-2.83} & \small{-3.8}\\
\small{$\Gamma$} & \small{300} & \small{1260} & \small{1000} \\
\small{$\alpha_{\nu}$} & \small{-1}  & \small{-0.17} & \small{0.8}\\
\small{$\beta_{\nu}$} & \small{-2}   & \small{-2.42} & \small{-2.39}\\
\small{$\gamma_{\nu}$} & \small{-4}   & \small{-4.42} & \small{-4.39}\\
\small{$\epsilon_{\nu\,1}$} & \small{$10^5$-$10^6$ GeV}  & \small{$1.2\times10^6$ GeV} & \small{$1.3\times10^6$ GeV} \\
\small{$\epsilon_{\nu\,2}$} & \small{~$10^7$ GeV}  & \small{$3.6\times10^8$ GeV} & \small{$5.6\times10^8$ GeV} \\
\small{z} & 1-2 & \small{0.903} & \small{1.822} \\
\small{$\delta t$} & \small{0.01 s} & \small{0.01 s*} & \small{0.053 s} \\
\small{$T_{90}$} & \small{2-1000 s} & \small{2.1 s} & \small{21.9 s} \\
\end{tabular*}
\end{center}
}
\end{table}

The values used to calculate the expected fireball neutrino rates from the FGST bursts, using only the Band component of the photon spectra, are listed in Table~\ref{tab:nuParms} together with the typical Waxman-Bahcall (WB) parameters. The numbers of neutrino events in IceCube estimated using the fireball formalism with the Band photon spectrum are reported in Table~\ref{tab:nuFire}.
\begin{table}[h!b!p!]
{
\caption{Fireball neutrinos in IceCube. The number reported here for GRB941017 is the sum of the number of events from each individual time bin. The parameters of the photon spectra for the time bins are taken from \citet{egret94} and parameters from Table~\ref{tab:nuParms} are used in the calculation of the number of neutrino events.}
\label{tab:nuFire}
}
{
\begin{center}
\begin{tabular*}{0.42\textwidth}{l|c}
 & \small{number of fireball $\nu$'s} \\[3pt]
\hline \hline
GRB941017 & $9.7\times10^{-2}$\\
GRB090510 & $1.6\times10^{-4}$\\
GRB090902b & $1.9\times10^{-3}$\\
\end{tabular*}
\end{center}
}
\end{table}

\section{Neutrinos from power-law GRB spectra}

\subsection{Fluences of secondary particles}

If we assume that the observed power-law components of the bursts
discussed here are due to the decay of neutral pions produced in
interactions between protons accelerated by shocks in the jet and
photons, we can predict the accompanying flux of muon neutrinos from
the production and decay of charged pions. The observed power-law
spectral components are quite flat ($E^{-1}-E^{-1.6}$), a spectral
behavior that is compatible with synchrotron radiation from highly
relativistic electrons and positrons which are produced when
high-energy photons scatter in the fireball photon field. 
For interactions of protons accelerated in shocks with fireball photons, we assume that the final neutrino spectrum will be of the form of Eq.~7. If we find the charged pion spectrum that, upon decay, produces the correct neutrino spectrum, we can determine the corresponding neutral pion spectrum and thus the gamma-ray spectrum. Assuming that the cascading process conserves energy, we can normalize the fluence of gamma-rays to the fluence of the measured power-law tail assuming its extension to some maximum energy $E_{\rm{max}}$
\begin{equation}
\label{eq:gamcasc}
\int_{0}^{10^{19}\,\rm{eV}}\!\! E_{\gamma}\, \frac{dN_{\gamma}}{dE_{\gamma}} \,dE_{\gamma} = \int_{E_{\rm{min}}}^{E_{\rm{max}}} \!\!E_{\gamma}\, \frac{dN^{\rm{HE}}_{\gamma}}{dE_{\gamma}}\,dE_{\gamma},
\end{equation}
where $E_{\rm{min}}$ is the minimum measured photon energy and
$E_{\rm{max}}$ is the proposed upper limit of the measured power-law
photon spectrum $dN^{\rm{HE}}_{\gamma}/dE_{\gamma}$. The upper limit of
$10^{19}$\,eV for the uncascaded gamma-ray spectrum assumes that the
parent protons extend to $\sim10^{20}$\,eV and that gamma-rays take on
average 1/10 of the parent proton energy. Due to the flatness of the
measured power-law spectra the precise value of $E_{\rm{min}}$ is
unimportant and the energy going into hadronic gamma rays is
determined purely by   $E_{\rm{max}}$. Since we assume that the
fireball is transparent for photons of energy $E_{\rm{max}}$, the
boost factor $\Gamma$ will vary as we vary the total energy in
hadronic photons. 

The algorithm for finding the flux of neutrinos is as follows. For a given $E_{\rm{max}}$, the measured parameters of the GRB will determine the minimum boost factor $\Gamma_{\rm{min}}$ (Eq.~5). These parameters and boost factor determine the neutrino spectral indices and break energies (Eqs.~8-10), giving us the unnormalized neutrino spectrum. Assuming that the neutrinos take on average 0.25 of the pion energy, we define the charged pion spectrum to be
\begin{equation}
\label{eq:pi+spec}
\frac{dN_{\pi^{+}}}{dE_{\pi}} = f_{\pi^{+}}\cdot
\begin{cases} 
\epsilon_{1,\pi}^{\alpha_{\nu}-\beta_{\nu}}E^{-\alpha_{\nu}}_{\pi}  & \mbox{if }E_{\pi}<\epsilon_{1,\pi}  \\
E^{-\beta_{\nu}}_{\pi} & \mbox{if }\epsilon_{1,\pi}<E_{\pi}<\epsilon_{2,\pi} \\
\epsilon_{2,\pi}^{\gamma_{\nu}-\beta_{\nu}}E^{-\gamma_{\nu}}_{\pi}  & \mbox{if }E_{\pi}>\epsilon_{2,\pi}  \\
\end{cases}
\end{equation}
with pion break energies $\epsilon_{i,\pi}$ relating to the neutrino break energies $\epsilon_{i}$ (Eqs.~\ref{eq:e1b} and\,\ref{eq:e2b}) via $\epsilon_{i,\pi} = 4\epsilon_{i}$. From the charged pion spectrum, we derive the neutral pion spectrum.
\begin{equation}
\label{eq:pi0spec}
\frac{dN_{\pi^{0}}}{dE_{\pi}} =2\cdot
\begin{cases} 
\dfrac{dN_{\pi^{+}}}{dE_{\pi}} & \mbox{if }E_{\pi}<\epsilon_{2,\pi}\\
(\epsilon_{2,\pi}^{\beta_{\nu}-\gamma_{\nu}}E^{-\beta_{\nu}+\gamma_{\nu}}_{\pi})\dfrac{dN_{\pi^{+}}}{dE_{\pi}}  & \mbox{if }E_{\pi}>\epsilon_{2,\pi}  \\
\end{cases}
\end{equation}
We assume that the photohadronic interaction takes place at the $\Delta$-resonance, giving twice as many neutral as charged pions. Moreover, the second break in the neutrino and charged pion spectra is due to the cooling of long-lived charged pions in the fireball and will not be present in the neutral pion spectrum. 

\begin{figure}[h]
\centering
\includegraphics[width=0.8\textwidth]{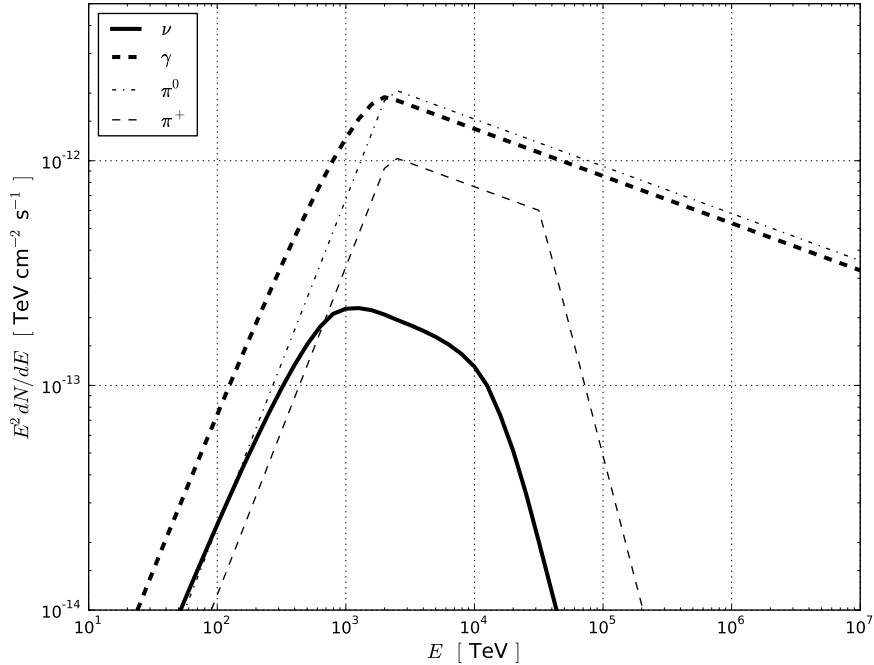}
\caption{Pion, neutrino and uncascaded gamma-ray spectra for Bin 2 of GRB941017, assuming a boost factor $\Gamma=300$. The spectra are not normalized but do account for the effect of neutrino oscillations.}
\label{fig:spectra}
\end{figure}

From the pion spectra we can find the decay spectra of gamma rays and neutrinos, normalized relative to the arbitrary factor of $ f_{\pi^{+}}$ in Eq.~\ref{eq:pi+spec} (see Fig.~\ref{fig:spectra}). Integrating over the neutrino and gamma-ray spectra then gives the amount of energy going into neutrinos relative to gamma rays.
\begin{equation}
\label{eq:gamnurelate}
\int_{0}^{5\times10^{18}\,\rm{eV}} \!\!E_{\nu}\, \frac{dN_{\nu}}{dE_{\nu}}\,dE_{\nu} = \eta\,  \int_{0}^{10^{19}\,\rm{eV}}\!\! E_{\gamma}\, \frac{dN_{\gamma}}{dE_{\gamma}} \,dE_{\gamma},
\end{equation}
The procedure for obtaining the decay particle spectra from the pion spectra Eqs.~\ref{eq:pi+spec} and\,\ref{eq:pi0spec} is described in Section 4.2 below.
 Finally, using Eq.~\ref{eq:gamcasc} with the factor $\eta$ found from Eq.~\ref{eq:gamnurelate}, we can find the actual amount of energy in the gamma-ray spectrum and therefore the absolute normalization of the neutrino flux:
\begin{equation}
\label{eq:henurelate}
\int_{0}^{5\times10^{18}\,\rm{eV}} \!\!E_{\nu}\, \frac{dN_{\nu}}{dE_{\nu}} \,dE_{\nu} = \eta\,   \int_{E_{\rm{min}}}^{E_{\rm{max}}} \!\!E_{\gamma}\, \frac{dN^{\rm{HE}}_{\gamma}}{dE_{\gamma}}\,dE_{\gamma}.
\end{equation}
Since neutrinos originating from pion decay have the flavor ratio $\nu_{e}:\nu_{\mu}:\nu_{\tau}=1:2:0$, over an astronomical baseline oscillations will transform the beam into the flavor ratio 1:1:1 if $\theta_{13}$ is small. Therefore the flux of muon-type neutrinos that reach the detector is half of the emitted flux and the final neutrino spectrum must be multiplied by a factor 0.5. 

In this method it is clear that higher observed maximum gamma-ray energies lead to more energy in neutrinos. This does not contradict the discussion in Section 2 that describes how a higher $E_{\rm{max}}$ implies a larger boost factor $\Gamma$ and hence a more transparent fireball. While a larger boost factor will imply a smaller photomeson production efficiency, relating $E_{\rm{max}}$ to the proton fraction $\epsilon_{p}/\epsilon_{e}$ via Figure~\ref{fig:gVsEmax941017} and Eq.~\ref{eq:epse} gives us $\epsilon_{p}/\epsilon_{e}\sim 10,000$ for $E_{\rm{max}}\sim 50$ GeV, to keep the total energy of the burst constant. A proton loading of this magnitude would likely compensate for the lowered photomeson production due to increased boost factor. The question of the baryonic loading is discussed in 
detail in Section \ref{eppe}. 

We also assume in this section that the observed power-law components correspond to all the energy emitted in hadronic gamma rays. There is, however, the possibility that some part of the high-energy gamma-ray tail will escape the GRB without interaction if the opacity is not large. These photons would cascade outside the source and are typically not observed\,\citep{razzaque2004,murase2008}. In that case, estimating the total hadronic gamma-ray energy using the observed photons would result in an underestimation of the energy and hence the neutrino flux. As a result, the neutrino fluxes derived in this section are from a minimum hadronic energy and are therefore conservative.  If GRBs are optically thin to photohadronic interactions, synchrotron self Compton \citep{guetta_granot2003,peer2004,stern2004} or proton synchrotron radiation \citep{dermer2, razzaque, asano} could dominate. Here, however, we posit that the sources are not optically thin to photohadronic interactions. It is clear that the observation of neutrinos with a full IceCube detector will resolve this issue.

\subsection{Pion decay spectra}

The muon neutrino and gamma-ray spectra are determined from pion spectra as follows: Both gamma rays are the product of the decay of the neutral pion $\pi^{0}\rightarrow \gamma + \gamma$. The two muon neutrinos are due to the decay chain $\pi^{\pm}\rightarrow \overset{\brabar} {\nu}_{\mu}  + \mu^{\pm}$ , $\mu^{\pm}\rightarrow \overset{\brabar} {\nu}_{\mu} + \overset{\brabar}{\nu}_{e} + e^{\pm}$. 

The gamma-rays and the first neutrino are found from standard two-body decay kinematics\,\citep{dermer,gaisser}:

\begin{equation}
\phi_{i}(E_{i}) = A_{i} \int_{E_{\pi\,\rm{min}}(E_{i})}^{\infty} \frac{\phi_{\pi}(E_{\pi})}{p_{\pi}}\,\,dE_{\pi}=A_{i} \int_{E_{\pi\,\rm{min}}(E_{i})}^{\infty} \frac{\phi_{\pi}(E_{\pi})}{\sqrt{E_{\pi}^{2} - m_{\pi}^{2}}}\,\,dE_{\pi}.
\end{equation}
For both particle types ($i=\nu,\gamma$), 
\begin{equation}
E_{\pi\,\rm{min}} = \frac{E_{i}}{1-r_{i}}+ (1-r_{i})\frac{m_{\pi}^{2}}{4E_{i}},
\end{equation}
where  $r_{\gamma}=0$ (neutral pion decay), $r_{\nu} = (m_{\mu}/m_{\pi})^{2}$ (charged pion decay), $A_{\gamma} = 2$, and $A_{\nu} = (1-r_{\nu})^{-1}$. Due to the $E + E^{-1}$ form of $E_{\rm{min}}$, the two-body spectra are symmetric (on a log-log scale) around a peak set by the pion mass and by the decay kinematics scale factor $r$. This peak is at $m_{\pi}/2 \simeq 70$ MeV for neutral pion decay and $(1-r_{\nu})m_{\pi}/2 \simeq 30$ MeV for charged pion decay. For a power-law distribution of pions with energies much larger than the pion mass the spectrum is a power-law of the same slope. 

The second neutrino is due to the decay of the muon from the charged pion $\mu^{\pm} \rightarrow e^{\pm} + \overset{\brabar} {\nu}_{e} + \overset{\brabar} {\nu}_{\mu}$. This is a three-body decay of a particle with a two-body decay energy distribution. The spectrum is\,\citep{gaisser}

\begin{equation}
\label{eq:mudec}
\phi_{\nu,2}(E_{\nu}) = \int^{\infty}_{E_{\mu\,\rm{min}}} dE_{\mu} \int^{E_{\pi\,\rm{max}}}_{E_{\pi\,\rm{min}}} \frac{dE_{\pi}}{(1-r_{\nu})} \frac{\phi_{\pi}(E_{\pi})}{\sqrt{E_{\pi}^{2}-m_{\pi}^{2}}} \frac{1}{E_{\mu}}\frac{dn}{dy}.
\end{equation}
The full limits on the integrals, valid at all energies, are
\begin{eqnarray*}
E_{\mu\,\rm{min}} &=& E_{\nu} + \frac{m_{\pi}^{2}}{4E_{\nu}}\\
E_{\pi\,\rm{max}} &=& \frac{E_{\mu}}{r_{\nu}} - (1-r_{\nu})\frac{m_{\pi}^{2}}{4E_{\mu}} \\
E_{\pi\,\rm{min}} &=& E_{\mu} + (1-r_{\nu})\frac{m_{\pi}^{2}}{4E_{\mu}}
\end{eqnarray*}
\citet{gaisser} considers only the spectrum at high energy and therefore omits the second term of each limit. The last term in the integrand $(1/E_{\mu})dn/dy$ is the muon decay distribution. It is given, for $y=E_{\nu}/E_{\mu}$, by

\begin{equation}
\frac{dn}{dy} = \frac{1}{\beta_{\mu}} \int^{x_{\rm{max}}}_{x_{\rm{min}}} \left[  f_{0}(x) \pm P_{\mu}f_{1}(x)\frac{2y-x}{\beta_{\mu}x} \right] \frac{dx}{x}
\end{equation}
where 
\begin{eqnarray*}
f_{0} &=& 2x^{2}(3-2x)\\
f_{1} &=& 2x^{2}(1-2x) \\
P_{\mu} &=& \frac{1}{\beta_{\mu}} \left( \frac{2E_{\pi}r_{\nu}}{E_{\mu}(1-r_{\nu})} - \frac{1+r_{\nu}}{1-r_{\nu}}   \right).
\end{eqnarray*}
$\pm P_{\mu}$ corresponds to the decay of $\mu^{\pm}$, respectively. The limits of the integral are 
\begin{eqnarray*}
x_{\rm{min}} &=& \frac{2y}{1+ \beta_{\mu}}\\
x_{\rm{max}} &=& \min[1,\frac{2y}{1- \beta_{\mu}}].
\end{eqnarray*}

As a practical matter, these spectra involve integrals to $+\infty$. This means that a change of variables is required before the integrals can be evaluated numerically. In this work we used the transform
\begin{equation}
\int^{\infty}_{a} f(x)\, dx = \int^{1/a}_{0} f\left(\frac{1}{t}\right) \frac{dt}{t^{2}}.
\end{equation}
We note in particular that the more general transform 
\begin{equation}
\label{eq:badtrans}
\int^{\infty}_{a} f(x)\, dx = \int^{1}_{0} f\left(a + \frac{1-t}{t}\right) \frac{dt}{t^{2}}
\end{equation}
does not evaluate the integrals for pion decay correctly around the low-energy kinematic peak. In the case of muon decay (Eq.~\ref{eq:mudec}) the integral transformed with Eq.~\ref{eq:badtrans} does not converge.

\subsection{Neutrino spectra and event rates}

\begin{figure}[htb]
\begin{center}
\includegraphics[width=0.8\textwidth]{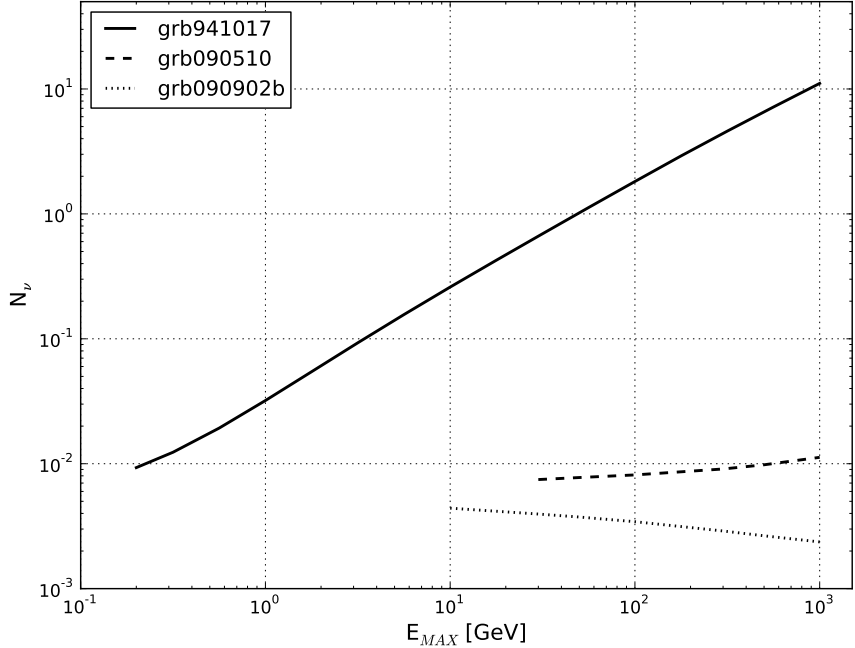}
\end{center}
\caption{Neutrino events in IceCube from 3 GRBs as a function of maximum observed gamma-ray energy.}
\label{fig:nuevents}
\end{figure}

Folding the derived neutrino fluxes with the IceCube effective area, we find the number of neutrino events as a function of neutrino energy and $E_{\rm{max}}$ of the observed gamma-ray fluence. In Fig.~\ref{fig:nuevents} we show the number of detected neutrinos with energy greater than detector threshold as a function of $E_{\rm{max}}$. We call attention to the fact that a burst with the same parameters as GRB941017, extending to  $E_{\rm{max}}>\!50$ GeV, will produce $>\!1$ neutrino event in IceCube. Given the absence of background events over such a short time interval, only a few events would be needed to constitute discovery of proton acceleration in GRBs. While very high $E_{\rm{max}}$ would require a (possibly) unphysically large boost factor and proton fraction, the detectability of a GRB941017-like burst with $E_{\rm{max}}$ within Fermi's energy range is highly encouraging.  

These neutrino event numbers are very different from the numbers derived with the standard fireball phenomenology using only the Band spectrum (Table~\ref{tab:nuFire}). While the two numbers of neutrino events for GRB941017 are similar if $E_{\rm{max}}\sim 200\,$MeV, the fireball event numbers for the Fermi bursts are both more than an order of magnitude smaller than the numbers calculated using the bolometric method, showing the large contribution of the GeV power-law component gamma rays relative to the lower-energy Band spectrum photons. 

%%%%%%%%%%%%%%%%%%%%%%%%%%%%%%%
\section{Photomeson production efficiency \label{eppe}}
%%%%%%%%%%%%%%%%%%%%%%%%%%%%%%%

The dependence of the pair-creation opacity $\tau_{\gamma\gamma}$ on the jet boost factor $\Gamma$ was presented in Eq.~\ref{eq:tau} where it was shown that $\tau_{\gamma\gamma}\propto \Gamma^{-4}$. The source must be therefore more transparent when photons with increasing energy $E_{\rm{max}}$ are detected. In addition to this, the efficiency of pion production, $f_{\pi}$, was shown to be proportional to $\tau_{\gamma\gamma}$~\citep{WB}. 
The predicted flux of neutrinos is therefore believed to be suppressed when $\Gamma$ becomes large. Here, we argue that this problem can be overcome by the requirement that GRBs are proton dominated. The parameter $\epsilon_p/\epsilon_e$ (an intrinsic burst parameter, hence a constant during the evolution of the burst) has to be $\gg 1$ to compensate for the decrease of the efficiency of pion production at large boost factors. However, since this is only the high-energy, nonthermal part of the energy balance, extremely high values of $\epsilon_p/\epsilon_e$ cannot be ruled out.\\

\begin{figure}[h!]
\begin{center}
\includegraphics[width=0.8\textwidth]{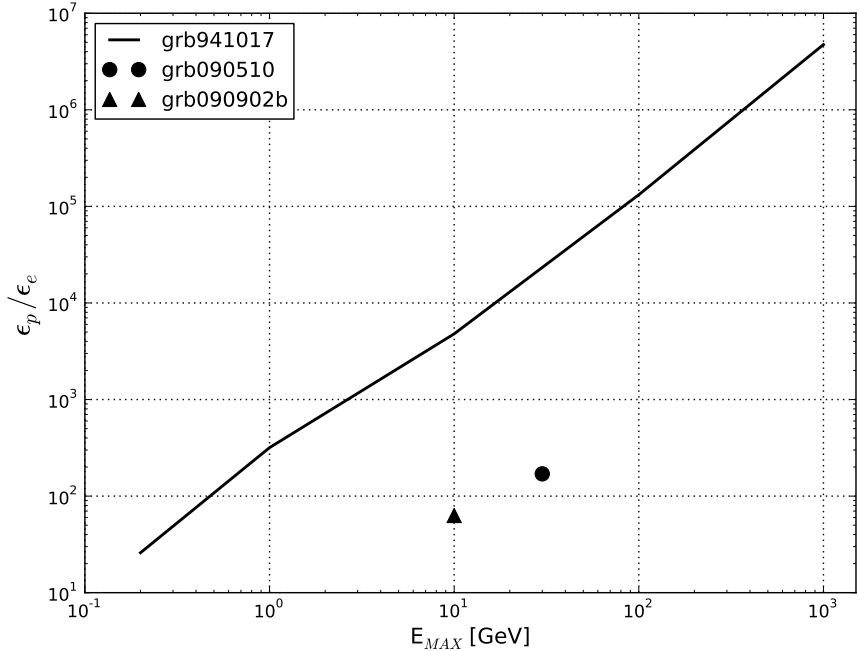}
\caption{GRB941017: $\epsilon_p/\epsilon_e$ Vs. observed $E_{\rm{max}}$.}
\label{fig:epee}
\end{center}
\end{figure}

\noindent In our discussion we have assumed that the high-energy component of all time bins of GRB941017 originates from the interaction of UHE CRs with photons of the spectral region described with the Band function. Here, we sum up the fluences of time bins 2-5 to get the total (time-integrated) fluence of the observed high-energy component:
\begin{equation}
F_{\mbox{\tiny{HE}}}^{\mbox{\tiny{TOT}}}(E_{\rm{max}})=\sum_{i=2}^5 F_{\mbox{\tiny{HE}}}^{i}(E_{\rm{max}})
\end{equation}
where:
\begin{equation}
F_{\mbox{\tiny{HE}}}^{i}(E_{\rm{max}})=\int_{30{\mbox{\tiny{\,keV}}}}^{E_{\rm{max}}} E_{\gamma} \frac{dN_{i,\gamma}}{dE_{\gamma}} dE_{\gamma}
\end{equation}
and $dN_{i,\gamma}/dE_{\gamma}$ is taken from Gonzalez et al.(2003). The index \textit{i} runs over time bins. The value of $E_{\rm{max}}$ is treated as a free parameter in all time bins since no break was detected by BATSE up to 200\,MeV. Therefore, we let it vary up to 1\,TeV and we calculate the corresponding observed total fluence of the high-energy power law. We then compare $F_{\mbox{\tiny{HE}}}^{\mbox{\tiny{TOT}}}(E_{\rm{max}})$, which is proportional to $\epsilon_p/\epsilon_e$:
\begin{equation}\label{eq:nu}
F_{\mbox{\tiny{HE}}}^{\mbox{\tiny{TOT}}}\propto \frac{\epsilon_p}{\epsilon_e}
\end{equation}
with the theoretical flux of photons from $\pi^0$-decay, $F_{\mbox{\tiny{HE}}}^{\mbox{\tiny{Theory}}}$.
In this way the hadronic model can reproduce the observed high-energy component for a given value of $\epsilon_p/\epsilon_e$. In Fig.~\ref{fig:epee} the ratio $\epsilon_p/\epsilon_e$ is plotted against $E_{\rm{max}}$. It is important to stress that the correct relation between $\Gamma$ and $E_{\rm{max}}$ has been used. We also note that for this GRB the values of $\epsilon_p/\epsilon_e$ shown in Fig.~\ref{fig:epee} refer to the time integrated fluence defined in Eq. 26. This is due to the assumption that $\epsilon_p/\epsilon_e$ is an intrinsic property of the burst and that therefore it should not vary with time.

\begin{figure}[h!]
\begin{center}
\includegraphics[width=0.8\textwidth]{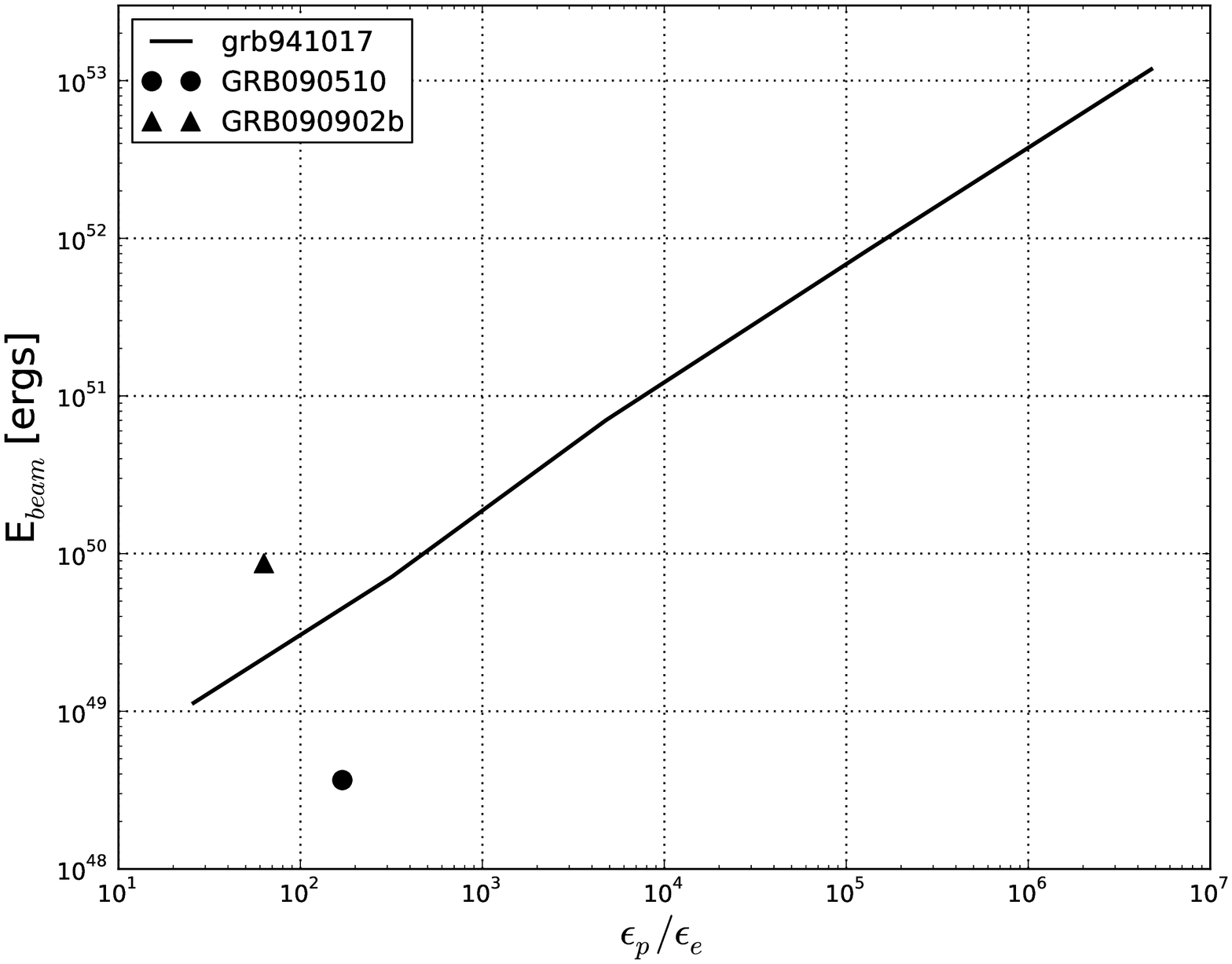}
\caption{GRB941017: beamed energy Vs. $\epsilon_p/\epsilon_e$}
\label{fig:beam}
\end{center}
\end{figure}

\noindent For each value of $\epsilon_p/\epsilon_e$ an estimate of the beamed energy release is then inferred using the relation:
\begin{equation}\label{eq:beam}
E_{\mbox{\tiny{beam}}}\approx \frac{\epsilon_p
  E_{\mbox{\tiny{TOT}}}^{\rm{iso}}}{2\Gamma^2}
\end{equation}
Results are shown in Fig.~\ref{fig:beam} and a physical limit can be inferred from the figure.
Very recently pair instability SNe have been shown to release $>10^{52}$\,ergs~\citep{galYam}. A similar energy release can be realistic in GRBs. Thus, we conclude that if beaming effects are considered the total energy release implied by the GRBs discussed here can be accomodated.\\

\noindent Applying the same argument to the time integrated spectra for GRB090510 and GRB090902b up to the observed $E_{\rm{max}}$ yields $\epsilon_p/\epsilon_e$\,$\sim$\,170 and $\sim$\,63 respectively and the beamed energy release using Eq.~\ref{eq:beam} with the corresponding numerical value of the jet boost factor (table~\ref{tab:nuParms}) is estimated to be $\sim$\,$3.66\times 10^{48}$\,ergs and $\sim$\,$8.70\times 10^{49}$\,ergs (Fig.~\ref{fig:beam}).

\section{Summary and Conclusions}
Three gamma-ray bursts with statistically significant high-energy
power-law spectral components have been detected thus far. 
While both leptonic and hadronic models have been proposed to explain the additional components,
  all models show some difficulties in explaining the observations. In this paper, we discuss the possibility of
  proton-dominated bursts with photohadronic interactions being
  responsible for the production of the high-energy component.
We calculate the associated fluxes of neutrinos for all three bursts.
First, the event rates of neutrinos are derived from standard fireball
phenomenology assuming that the total energy in protons is 10 times more than in electrons. In a second approach, we take the existence of the
high-energy power-law components as indicative of the decay of
$\pi^{0}$-mesons and keep the ratio of the energy going into
non-thermal electrons and protons as a free parameter. 
This allows us to calculate the magnitude of the neutrino flux from
the related charged pions,
and we find that a burst like GRB941017 
will produce at least one neutrino event and hence be detectable by
IceCube if its power-law component extends to energies in excess of
$\sim50$\,GeV. While the event rates for the Fermi GST bursts are small, this is due to their large boost factors and redshifts as opposed to a burst like GRB941017, which is nearby with low boost factor.  IceCube and the future observatory Km3NeT will be able
to help determine if the high-energy components in GRBs are indeed due
to photohadronic interactions or if other scenarios are more viable.

\acknowledgments
J.K.B. and M.O. acknowledge support from the Research Department of Plasmas with Complex Interactions (Bochum). F.H. and A.\'O.M were supported in part by the National Science Foundation under Grant No.~OPP-0236449 and in part by the University of Wisconsin Alumni Research Foundation.

%===========BIBLIOGRAPHY===================%


\begin{thebibliography}{}

\bibitem[Abbasi\,\textit{et al.}(2009) ]{kappes}
Abbasi, R. \textit{et al.}, \apj \, \textbf{701}, 1721 (2009)

\bibitem[Abdo\,\textit{et al.}(2009) ]{090510}
Abdo, A.A. \textit{et al.}, arXiv.0908.1832 (2009)

\bibitem[Ahlers\,\textit{et al.}(2005) ]{ahlers}
{Ahlers}, M. \textit{et al.}, \prd\, \textbf{72}, 023001(2005)

\bibitem[Alvarez-Mu\~{n}iz\,\textit{et al.}(2002) ]{bolo}
Alvarez-Mu\~{n}iz, J. \textit{et al.}, \apj \, \textbf{576}, L33 (2002)

\bibitem[Alvarez-Mu\~{n}iz\,\textit{et al.}(2004) ]{hh_04}
Alvarez-Mu\~{n}iz, J. \textit{et al.}, \apjl \, \textbf{604}, L85 (2004)

\bibitem[Asano\,\textit{et al.}(2009) ]{asano}
Asano, K. \textit{et al.}, \apjl \, \textbf{705}, L191 (2009)

\bibitem[Asano\,\textit{et al.}(2009) ]{grbrate}
Asano, K. \textit{et al.}, \apj \, \textbf{699}, 953 (2009)

\bibitem[Band\,\textit{et al.}(1993) ]{band}
Band, D. \textit{et al.}, \apj\, \textbf{413}, 281 (1993)

\bibitem[Becker\,\textit{et al.}(2006) ]{becker}
Becker, J.K. \textit{et al.}, Astroparticle Physics \,\textbf{25}, 118 (2006)

\bibitem[Bissaldi\,\textit{et al.}(2009) ]{090902b}
Bissaldi, E. \textit{et al.}, arXiv:0909.2470 (2009)

\bibitem[Dermer(1986) ]{dermer}
Dermer, C., \apj \, \textbf{307}, 47 (1986)

\bibitem[Dermer\,\textit{et al.}(2004) ]{dermer2}
Dermer, C. and Atoyan, A., \aap \, \textbf{418}, L5 (2004)

\bibitem[Gaisser(1990) ]{gaisser}
Gaisser, T., \textit{Cosmic Rays and Particle Physics} (Cambridge University Press, Cambridge, England 1990)

\bibitem[A. Gal-Yam\,\textit{et al.}(2009) ]{galYam}
{Gal-Yam, A. \textit{et al.}, \nat\, \textbf{462}, 8579 (2009)}

\bibitem[Gonz{\'a}lez\,\textit{et al.}(2003) ]{egret94}
{Gonz{\'a}lez}, M. M. \textit{et al.}, \nat\, \textbf{424}, 7491 (2003)

\bibitem[Gonz{'a}lez-Garcia\,\textit{et al.}(2009) ]{effarea}
Gonzalez-Garcia, M.C. \textit{et al.}, Astroparticle Physics \,\textbf{31}, 437 (2009)

\bibitem[Gould\,\textit{et al.}(1967) ]{gould}
Gould, R. J. and Schr\'eder, G. P., Phys. Rev. \, \textbf{155}, 1404 (1967)

\bibitem[Granot \textit{et al.} (2003) ]{guetta_granot2003}
Granot, J. and Guetta, D., ApJL \textbf{598}, 11 (2003)

\bibitem[Guetta\,\textit{et al.}(2004) ]{guetta}
Guetta, D. {\it et al.}, Astroparticle Physics \textbf{20}, 429 (2004)

\bibitem[Lithwick\,\textit{et al.}(2001) ]{Gmin}
Lithwick, Y. and Sari, R., \apj\, \textbf{555}, 540 (2001)

\bibitem[Murase\,\textit{et al.}(2006) ]{murase}
Murase, K. and Nagataki, S., \prd\, \textbf{73}, 063002 (2006)

\bibitem[Murase\,\textit{et al.}(2008) ]{murase2008}
Murase, K. {\it et al.}, \prd\, \textbf{78}, 023005 (2008)

\bibitem[Murase\,\textit{et al.}(2009) ]{murase2009}
Murase, K. {\it et al.}, \prl\, \textbf{103}, 081102 (2009)

\bibitem[Pe'er\,\textit{et al.}(2004) ]{peer2004}
Pe'er, A. and Waxman, E., ApJL\, \textbf{603}, L1 (2004)

\bibitem[Razzaque\,\textit{et al.}(2004) ]{razzaque2004}
Razzaque, S.  {\it et al.}, \apj\, \textbf{613}, 1072 (2004)

\bibitem[Razzaque\,\textit{et al.}(2009) ]{razzaque}
Razzaque, S.  {\it et al.},  arXiv:0908.0513 (2009)

\bibitem[Rachen\,\textit{et al.}(1998) ]{rachen}
Rachen, J.P. and Me\'sza\'ros, P., \prd\, \textbf{58}, 123005 (1998)

\bibitem[Stern \,\textit{et al.}(2004) ]{stern2004}
Pe'er, A. and Waxman, E., MNRAS\, \textbf{352}, L35 (2004)

\bibitem[Vietri(1995) ]{vietri_1995}
Vietri, M., \apj\, \textbf{453}, 883 (2005)

\bibitem[Waxman(1995) ]{waxman_1995}
Waxman, E., \prl\, \textbf{75}, 386 (1995)

\bibitem[Waxman(2004) ]{waxman2004}
Waxman, E., \apj\, \textbf{606}, 988 (2004)

\bibitem[Waxman\,\textit{et al.}(1997) ]{WB}
Waxman, E. and Bahcall, J., \prl\, \textbf{78}, 2292 (1997)


\end{thebibliography}
\end{document}